\documentclass[9pt,
prd,
amsmath,
amssymb,
preprint,
author-year,
showpacs,
showkeys,
a4paper
]{revtex4-1}

\usepackage{dcolumn}
\usepackage{bm}
\usepackage{amsmath,amssymb,fancybox,amscd,accents,mathptmx}
\usepackage[all]{xy}
\usepackage{braket}
\usepackage{makeidx}
\usepackage{mathrsfs}
\usepackage{graphicx}

\newcommand{\dis}{\displaystyle}

\newcommand{\pa}{\partial}

\newcommand{\lam}{\lambda}

\newcommand{\ep}{\epsilon}

\newcommand{\al}{\alpha}

\newcommand{\no}{\nonumber}

\def\bee{\begin{equation}}
\def\ede{\end{equation}}
\def\bna{\begin{eqnarray}}
\def\ena{\end{eqnarray}}
\def\bgen{\begin{enumerate}}
\def\eden{\end{enumerate}}
\def\bede{\begin{description}}
\def\ende{\end{description}}
\def\bex{\begin{axio}}
\def\enx{\end{axio}}
\def\bef{\begin{defi}}
\def\enf{\end{defi}}
\def\bepp{\begin{prop}}
\def\enpp{\end{prop}}
\def\bet{\begin{theo}}
\def\ent{\end{theo}}
\def\bel{\begin{lemm}}
\def\enl{\end{lemm}}
\def\bep{\begin{proo}}
\def\enp{\end{proo}}
\def\bepb{\begin{prob}}
\def\enpb{\end{prob}}
\def\bpri{\begin{prin}}
\def\epri{\end{prin}}

\begin{document}
\preprint{KOBE-COSMO-17-06}
\title{Electromagnetic Memory Effect Induced by Axion Dark Matter}

\author{Daiske Yoshida}
 \email{dice-k.yoshida@stu.kobe-u.ac.jp}
 \affiliation{Department of Physics, Kobe University, Kobe, 657-8501, Japan}
 
\author{Jiro Soda}
 \email{jiro@phys.sci.kobe-u.ac.jp}
 \affiliation{Department of Physics, Kobe University, Kobe, 657-8501, Japan}
 

\date{\today}


\begin{abstract}
Memory effects of gravitational waves from astronomical events or a primordial universe might have the information of new physics. It is intriguing to observe that the memory effect exists in electrodynamics as a net momentum kick, while the memory effect in gravity appears as a net relative displacement. In particular, Winicour has shown that the B-mode memory, which characterizes parity-odd global distribution of memory, does not exist. We study the memory effect in axion electrodynamics and find that the B-mode memory effect can exist, provided the existence of coherently oscillating axion background field. Moreover, we examine the detectability of the axion dark matter using this effect.
\end{abstract}


\maketitle

\section{Introduction}

Recently, gravitational memory effect~\cite{ZelPol:1974} 
has been intensively discussed in conjunction with 
fundamental aspects of gravity \cite{0264-9381-33-17-175006,0264-9381-29-21-215003,Braginsky:1987aa,PhysRevD.89.084039} and observations \cite{0004-637X-752-1-54,PhysRevLett.117.061102,PhysRevLett.67.1486}. 
The memory effect is relevant to many issues such as the information problem of black holes \cite{PhysRevLett.116.231301}, Bondi-Metzner- Sachs (BMS) translation \cite{Strominger2016}, cosmology \cite{PhysRevD.94.044009,1475-7516-2016-05-059} and others \cite{Bieri:2010tq}. 
Moreover, the memory of the scalar gravitational waves of the scalar-tensor theory was discussed in \cite{Du:2016hww}.

Interstingly, there exists an electromagnetic counter part of the memory effects~\cite{0264-9381-31-20-205003,0264-9381-30-19-195009}.
In particular, Winicour studied the global aspect of memory effect, that is, the E-mode and B-mode memory effects, 
in conventional electrodynamics \cite{0264-9381-31-20-205003}. 
He found that the B-mode memory can not be realized by physically realistic sources.
Furthermore, M$\ddot{\rm a}$dler and Winicour extended their analysis to the E-mode and B-mode memory effects of gravitational waves \cite{0264-9381-33-17-175006}.

As the electromagnetic waves have been used to probe the Universe, it is reasonable to inquire if we can utilize electromagnetic memory to investigate the cosmological issues. In this paper, we focus on the dark matter problem. 
As is well known, the axion that is coherently oscillating on the halo scales is one of the candidates for the dark matter. Here, therefore, we assume the axion dark matter dominates the dark matter component.
The point is that the axion modifies the conventional electrodynamics and hence affects the electromagnetic memory effect. 
Moreover, since the presence of the axion violates the parity symmetry, it is interesting to see if the axion dark matter can induce the B-mode electromagnetic memory or not.
Therefore, by extending the work in Ref. \cite{0264-9381-31-20-205003}, we study the global aspects of memory effect in axion electrodynamics. 
We show that the axion dark matter indeed leads to the parity-violating electromagnetic memory, namely, the B-mode memory.

The organization of this paper is as follows. 
In Sec. II, we shortly review the axion electrodynamics.
In Sec. III, we give the settings and equations of motion of axion electrodynamics in a null retarded coordinate system.
In Sec. IV, we show the nontrivial B-mode memory induced by the axion field. We also discuss its detectability.
The final section is devoted to the conclusion.

\section{Axion Electrodynamics}

In this section, we define axion electrodynamics, which is the electrodynamics with an axion coupled to the electromagnetic fields. It is well known that the axion is a strong candidate for the dark matter.
The axion has been well studied because it solves $\theta$ problem in QCD~\cite{Peccei:1977hh}.
Moreover, it is widely recognized that axions are ubiquitous in string theory~\cite{Svrcek:2006yi,Arvanitaki:2009fg}. 


The action for 4-dimensional axion electrodynamics is given by
\bee
S = \int dx^4 \left[ \hspace{2mm} - \frac{1}{4} F_{\mu\nu} F^{\mu\nu} + A_{\mu}J^{\mu} 
       -\frac{1}{2} \partial_{\mu}\Phi \, \partial^{\mu}\Phi - U(\Phi) - \frac{\lam}{4 } \Phi F_{\mu\nu} \tilde{F}^{\mu\nu} \right] 
 \ , 
\ede
where $\lambda$ is a coupling constant, $J^\mu$ is a charge current, $U(\Phi)$ is a potential function for an axion field $\Phi$, and  $A_{\mu}$ is a gauge field with the field strength defined by
$
F_{\mu\nu} = \partial_{\mu} A_{\nu} - \partial_{\nu} A_{\mu} \ .
$
Note that the field strength satisfies the identity
$
\partial_{[\mu}F_{\mu\nu]} = 0 \ .
$
The dual of the field strength $\tilde{F}^{\mu\nu}$ is defined by
\bee
\tilde{F}^{\mu\nu}= \frac{1}{2} \ep^{\mu\nu\al\beta}F_{\al\beta}.
\ede
The variation of the action with respect to the gauge field gives equations of motion for the gauge field,
\bee
\partial_{\mu} F^{\al \mu} = \dis J^{\al} - \frac{\lam}{2} \ep^{\al\mu\nu\lam} \partial_{\mu}\left( \Phi F_{\nu\lam} \right)  .
\ede 
Here, we introduced a charge current satisfying the conservation law,
\bee
\partial_{\mu} J^{\mu} = 0.
\ede 
The equation of motion for the axion is given by
\bee
\partial_{\mu} \partial^{\mu} \Phi - \pa_{\Phi} U(\Phi) = \frac{\lam}{4} F_{\mu\nu} \tilde{F}^{\mu\nu} . \label{SEoM}
\ede
For simplicity, we consider the potential 
\bee
 U(\Phi) = \frac{1}{2}m^{2} \Phi(x)^2,
\ede
where $m$ is the mass of the axion. 
We will focus on the homogeneous oscillating solution for the axion because the massive radiation mode of scalar field decays at null infinity \cite{Winicour:1988aq,Helfer:1993hv}.

\section{Basic equations in retarded null coordinate system}

In this section, we derive basic equations for analyzing the memory effect in axion electrodynamics. We employ the perturbation method with a coupling constant as a small parameter. We also decompose the gauge potential into parity-even and -odd parts.


It is convenient to use the retarded coordinate system for describing the memory effect in Ref. \cite{0264-9381-31-20-205003}. 
By using the spherical coordinate, $(t,r,\theta ,\phi)$, the retarded time $u$ is defined by
\bee
u = t-r \ .
\ede
So, the metric in the retarded coordinate system is given by 
\bna
ds^2 &=& g_{\mu\nu} {\rm d}x^{\mu} \,{\rm d}x^{\nu} \no \\
	 &=& - {\rm d}u^2 - 2 {\rm d}u \, {\rm d}r + r^2 q_{AB}\, {\rm d}x^{A} \,{\rm d}x^{B},
\ena
where $q_{AB}$ is the metric of a two-dimensional sphere. The derivatives in both coordinate systems are related by
\bee
\left\{ \begin{array}{l} \dis \pa_u = \pa_t, \\ \dis \pa_r = \pa_t + n^i \pa_i \hspace{3mm} {\rm and} \\ \dis \pa_A = r \left( \pa_A n^i \right)\pa_i. \end{array} \right.
\ede
The right-hand side is the derivative in the Cartesian coordinate system, and $n^i$ is a unit radial vector defined by 
\bee
n^i = \frac{x^i}{r} = (\sin{\theta} \cos{\phi} , \sin{\theta} \sin{\phi},\cos{\theta}) \, \, \, .
\ede
We use the gauge freedom to set 
\bee
A_r = 0 \ .
\ede
So, we have the formulas for the field strength as
\bee
\left\{ \begin{array}{l} 
\dis F_{ru} = \pa_r A_u = n^i F_{it}, \\
\dis F_{rB} = \pa_{r} A_{B} = r \left( \pa_{B}n^{j} \right) \left( F_{tj} + n^{i} F_{ij} \right), \\
\dis F_{Bu} = \pa_{B} A_{u} - \pa_{u}A_{B} = r \left( \pa_{B} n^{i} \right) F_{it} \hspace{3mm}{\rm and}\\
\dis F_{BC} = \pa_{B} A_{C} - \pa_{C} A_{B} = r^{2} \left( \pa_{B} n^{i} \right) \left( \pa_{C} n^{j} \right) F_{ij} .
\end{array} \right.
\ede
These relations are useful in evaluating $r$-dependence of physical fields.
Note that the gauge potential of angular direction, $A_{B}$, can be represented by
\bee
A_{B} = \al_{:B} + \ep_{BC} \beta^{:C},
\ede
where the colon denotes a covariant derivative and $\ep_{BC}$ is the Levi-Civit\`a tensor on a two-dimensional sphere. 
The variables $\al$ and $\beta$ correspond to the E-mode and B-mode radiations, respectively.


Now, we investigate the memory effect in axion electrodynamics. Since the equations of motion are nonlinear,
we have to resort to the perturbation method with a coupling constant $\lam $ as the parameter.
Then, we can expand the gauge field as
\bee
A_{\al} = A^{(0)}_{\al} + \lambda A^{(1)}_{\al} + \cdots .
\ede
Similar expansion applies to the axion field.


If we give the source of electrical field of a charged point particle, 
the electromagnetic field and the axion field at the lowest order can be separately solved.
In the retarded null coordinate system, the equation of motion for the axion field $(\ref{SEoM})$ is
\bee
\left( \pa^{2}_{r}-2 \pa_{u}\pa_{r} -\frac{2}{r}\pa_{u} + \frac{2}{r}\pa_{r} \right)\Phi^{(0)} + \frac{1}{r^{2}}\Phi^{(0)\,:A}_{\,\,\,\,:A} - m^{2} \Phi^{(0)} = 0.
\ede
Thus, the oscillating solution for the axion is given by
\bee
\Phi^{(0)}(x) = D \cos{\left\{ m \left( u+r \right) + \rho \right\}} ,
\ede
where $D$ and $\rho$ are arbitrary constants. For simplicity, we set the phase $\rho$ to be zero.

\begin{widetext}

Now, the equations of motion at the lowest order read
\bna
&& \hspace{5mm} \pa_{r} \left( r^{2} \pa_{r}A^{(0)}_{u} \right) - \left( \pa_{r} \alpha^{(0)} \right)^{:B}_{\,\,\,\,:B} = r^{2} J^{u}, \\
&& \hspace{5mm} - r^{2} \pa_{u} \pa_{r} A^{(0)}_{u} + \left( \pa_{r} \alpha^{(0)} \right)^{:B}_{\,\,\,\,:B} - \left( \pa_{u}\alpha^{(0)} - A^{(0)}_{u} \right)^{:B}_{\,\,\,\,:B} = r^{2} J^{r}, \\
&& \hspace{5mm} \pa_{r} \left( 2\pa_{u} - \pa_{r} \right) \left( \alpha^{(0)} \right)^{:B}_{\,\,\,\,:B} - \left( \pa_{r} A^{(0)}_{u} \right)^{:B}_{\,\,\,\,:B} = r^{2} J^{B}_{\,\,\,\,:B} \hspace{5mm} {\rm and}\\
&& \hspace{5mm} \pa_{r} \left( 2\pa_{u} - \pa_{r} \right) \left( \beta^{(0)} \right)^{:B}_{\,\,\,\,:B} - \frac{1}{r^{2}}\left( \beta^{(0)} \right)^{:C\,\,:B}_{\,\,\,\,:C\,\,:B} = r^{2} \ep_{BC}J^{B:C}.
\ena
The above equations are the same as those in Ref. \cite{0264-9381-31-20-205003}. 
The difference appears at the next-order equations as follows:
\bna
&& \hspace{5mm} \pa_{r} \left( r^{2} \pa_{r}A^{(1)}_{u} \right) - \left( \pa_{r} \alpha^{(1)} \right)^{:B}_{\,\,\,\,:B} =  \left( \pa_{r}\Phi \right) \left( \beta^{(0)} \right)^{:C}_{\,\,\,\,:C} \label{puF}, \\
&& \hspace{5mm} - r^{2} \pa_{u} \pa_{r} A^{(1)}_{u} + \left( \pa_{r} \alpha^{(1)} \right)^{:B}_{\,\,\,\,:B} - \left( \pa_{u}\alpha^{(1)} - A^{(1)}_{u} \right)^{:B}_{\,\,\,\,:B} = - \left( \pa_{u}\Phi \right) \left( \beta^{(0)} \right)^{:B}_{\,\,\,\,:B} \label{prF} ,\\ 
&& \hspace{5mm} \pa_{r} \left( 2\pa_{u} - \pa_{r} \right) \left( \alpha^{(1)} \right)^{:B}_{\,\,\,\,:B} - \left( \pa_{r} A^{(1)}_{u} \right)^{:B}_{\,\,\,\,:B} = \left( \pa_{u} \Phi \right) \left( \pa_{r}\beta^{(0)} \right)^{:B}_{\,\,\,\,:B} - \left( \pa_{r} \Phi \right) \left( \pa_{u}\beta^{(0)} \right)^{:B}_{\,\,\,\,:B}  \label{peF} \hspace{3mm}{\rm and} \\ 
&& \hspace{5mm} \pa_{r} \left( 2\pa_{u} - \pa_{r} \right) \left( \beta^{(1)} \right)^{:B}_{\,\,\,\,:B} - \frac{1}{r^{2}}\left( \beta^{(1)} \right)^{:C\,\,:B}_{\,\,\,\,:C\,\,:B}  = - \left( \pa_{u}\Phi \right)\left( \pa_{r} \al^{(0)} \right)^{:B}_{\,\,\,\,:B} + \left( \pa_{r} \Phi \right) \left( \pa_{u}\al^{(0)} - A_{u}^{(0)} \right)^{:B}_{\,\,\,\,:B} \label{pbF}.\hspace{1cm}
\ena
\end{widetext}
Here, we set the four-dimensional Levi-Civat\`a tensor as
\bee
\epsilon^{\mu\nu\rho\sigma} \equiv \frac{1}{\sqrt{-g}}\tilde{\epsilon}^{\mu\nu\rho\sigma} \hspace{3mm} {\rm and} \hspace{3mm} \tilde{\epsilon}^{u r \theta \phi} \equiv 1,
\ede
and rewrite components as follows:
\bna
\epsilon^{urAB} &=& \frac{1}{r^{2}}\frac{1}{\sqrt{q}}\tilde{\epsilon}^{AB} \no \\
								&=& \frac{1}{r^{2}}\epsilon^{AB} .  
\ena
We see the background fields act as the sources of the first-order fields. In fact, the axion mediates the the conversion 
from the E-mode memory to the B-mode memory.
In the next section, we will give a simple example 
for the B-mode radiation memory, which is generated by the axion.

\section{Memory Effect in Axion Electrodynamics}

We are now in a position to discuss memory effect in axion electrodynamics.
First, we derive the E-mode memory effect induced by a charged particle.
Second, we show the B-mode memory effect appears due to the axion coupling. 


\subsection{E-mode Memory Effect}

At null infinity, the angular component of electric field, $E_{B} = F_{Bu}$, can survive as the $\mathcal{O}(1)$ quantity in the asymptotic limit $r\rightarrow \infty$~\cite{0264-9381-31-20-205003,0264-9381-30-19-195009}. 
Hence, the memory effect can be defined by
\bna
\int^{\infty}_{- \infty}\, du \, E_{B} = \int^{\infty}_{- \infty}\, du\, \left( A_{u} - \pa_{u}\al \right)_{:B} \, du 
  - \ep_{BC} \int^{\infty}_{- \infty}\, \pa_{u}\beta^{:C} \, du \, \, \, .
\ena
On the right-hand side, the first term gives the E-mode radiation memory, and the second term gives the B-mode radiation memory. These components can be extracted by taking the divergence or curl of two-dimensional metric. 
Indeed, we obtain the formula for the E-mode memory,
\bee
\dis \int^{\infty}_{- \infty}\, du \, E_{B}^{\,\,\,\,:B} = \int^{\infty}_{- \infty}\, du \, \left( A_{u} - \pa_{u}\al \right)_{:B}^{\,\,\,\,:B} \,\,\, ,
\ede
and that for the B-mode memory,
\bee
\int^{\infty}_{- \infty}\, du \, \epsilon^{BC} E_{B\, :C} = - \int^{\infty}_{- \infty}\, du \,  \left( \pa_{u} \beta\right)_{:B}^{\,\,\,\,:B}\,\,\, .
\ede
So, we must evaluate these quantities by using the equations of motion for fields.
In this section, we will give the memory effect in axion electrodynamics
with the perturbation method.


We consider a charged particle with a current,
\bna
J^{t}\equiv\rho(t,{\bf x}) &=& q \delta^{(3)}({\bf r} - {\bf r}(t)) \hspace{3mm} {\rm and}\\
J^{i}\equiv{\bf j}(t,{\bf x}) &=& q {\bf v}(t)\delta^{(3)}({\bf r} - {\bf r}(t))\,\,\, ,
\ena
where $\delta^{(3)}({\bf x})\equiv\delta(x)\,\delta(y)\,\delta(z)$
is a three-dimensional delta function. 
This source gives the Lienard-Wiechert potential in Lorentz gauge in pure Maxwell theory. If the charged particle is accelerated, it generates the electromagnetic radiation. Generally, the radiation part can survive at infinity $r\to \infty$. For simplicity, we assume that the particle moves to $+z$ direction and the velocity of this particle is smaller than the light of velocity. Then, the angular components of the electric field in the null coordinate system are given by
\bee
E_{\theta} \simeq \frac{q}{4 \pi}a(u) \sin{\theta} \hspace{5mm} {\text at}\hspace{5mm} r \to \infty
\ede
and
\bee
E_{\phi} = 0.
\ede
Here, $a(u)$ represents an acceleration of the charged particle.

\subsection{B-mode memory effect}
\begin{figure}[htbp]
  \begin{center}
        \begin{center}
          \includegraphics[clip, width=9cm,bb=0 0 1024 768]{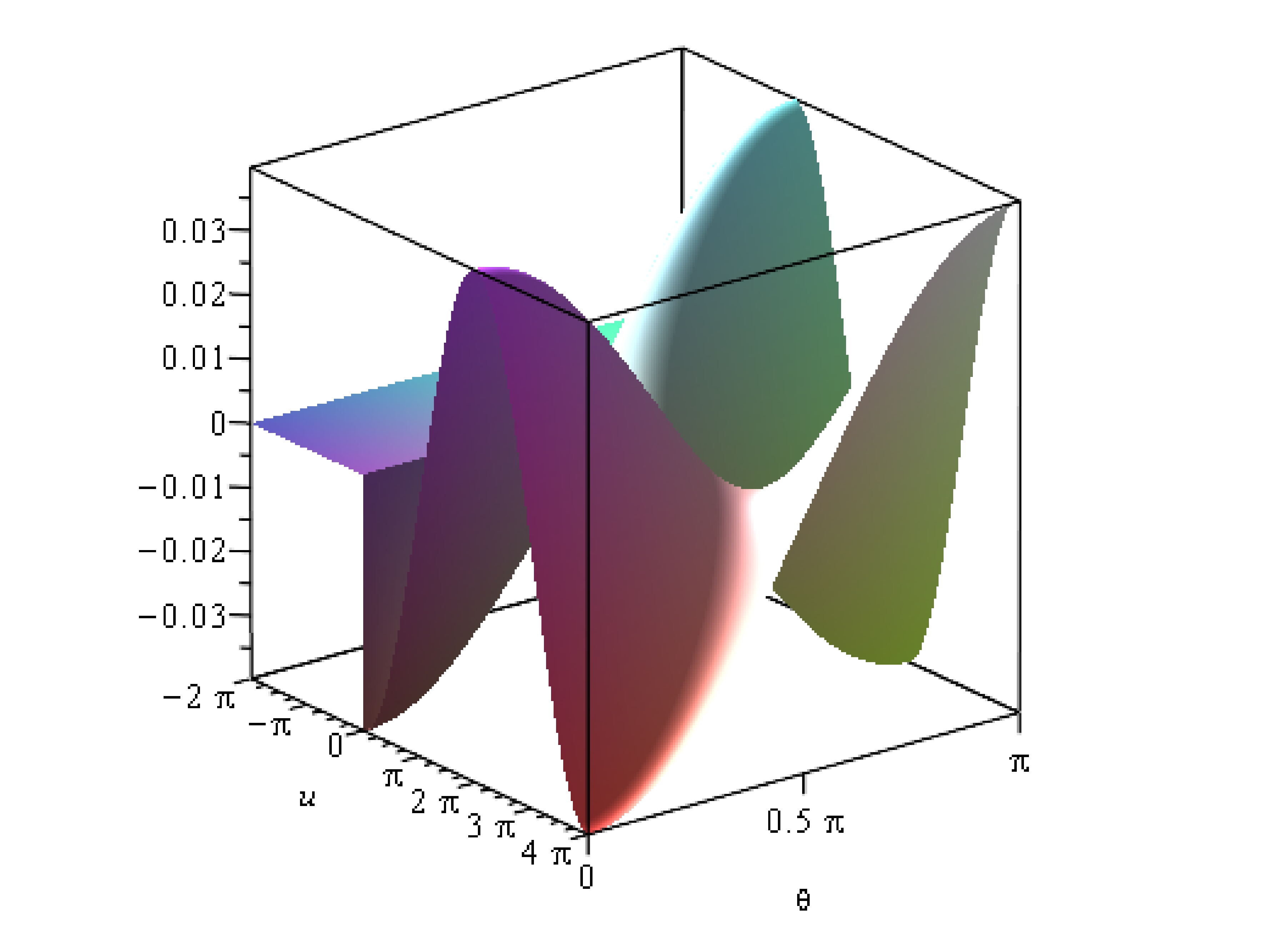}
        \end{center}
  \caption{The time evolution of perturbed B-mode memory ( $D=1,q=1,m=1,V=1/2$ )}
   \label{F.2}
  \end{center}
\end{figure}

Given the solution induced by a charged particle, we can easily calculate the correction to the memory 
represented by the perturbed gauge field $A^{(1)}_{\mu}$. 
First of all, since the source has no rotation part of the current, 
 the parity-odd part in the background vanishes,
\bee
\beta^{(0)}=0 \ .
\ede
Thus, there is no correction for the parity-even part at the first order through Eqs. $(\ref{puF})$, $(\ref{prF})$ and $(\ref{peF})$,
\bee
E_{B}^{(1)\,:B} = 0 \ .
\ede
However, there exist corrections to the parity-odd part. Notice that the equation of motion at $r \to \infty$ is given by
\bee
\pa_{r} \left( 2\pa_{u} - \pa_{r} \right) \left( \beta^{(1)} \right)^{:B}_{\,\,\,\,:B} = - \left( \pa_{r} \Phi \right) E_{B}^{(0)\,:B} \ .
\ede
Here, we neglected the second term on the left-hand side and the first term on the right-hand side of $(\ref{pbF})$.
Hence, the parity-even background electric field generates the correction to the parity-odd memory. 
We now derive the perturbed expression for the memory effect by the burst of a charged particle which is rest for $t < 0$
 and moves along z-axis with the constant velocity $V$ normalized by the velocity of light $c$ for $t\geq 0$ . 
The velocity is represented by
\bee
v(t) \equiv V \cdot H(t) \ ,
\ede
where we used the Heaviside step function
\bee
H(t) \equiv \left\{ \begin{array}{l} \hspace{5mm} 1 \hspace{1cm} (t\geq 0)\ , \\ 
\hspace{5mm} 0 \hspace{1cm} (t < 0) \ . 
\end{array} \right.
\ede
A derivative of the Heaviside function gives the delta function
\bee
\delta(t) \equiv \frac{d H(t)}{dt} \ .
\ede
Thus, the parity-even part of the electric field at the lowest order is given by
\bee
E_{B}^{(0) :B} = \frac{q V}{2 \pi}\delta ( u ) \cos{\theta} \ .
\ede
\begin{figure}[htbp]
 \begin{center}
        \begin{center}
          \includegraphics[clip, width=9cm,bb=0 0 1024 768]{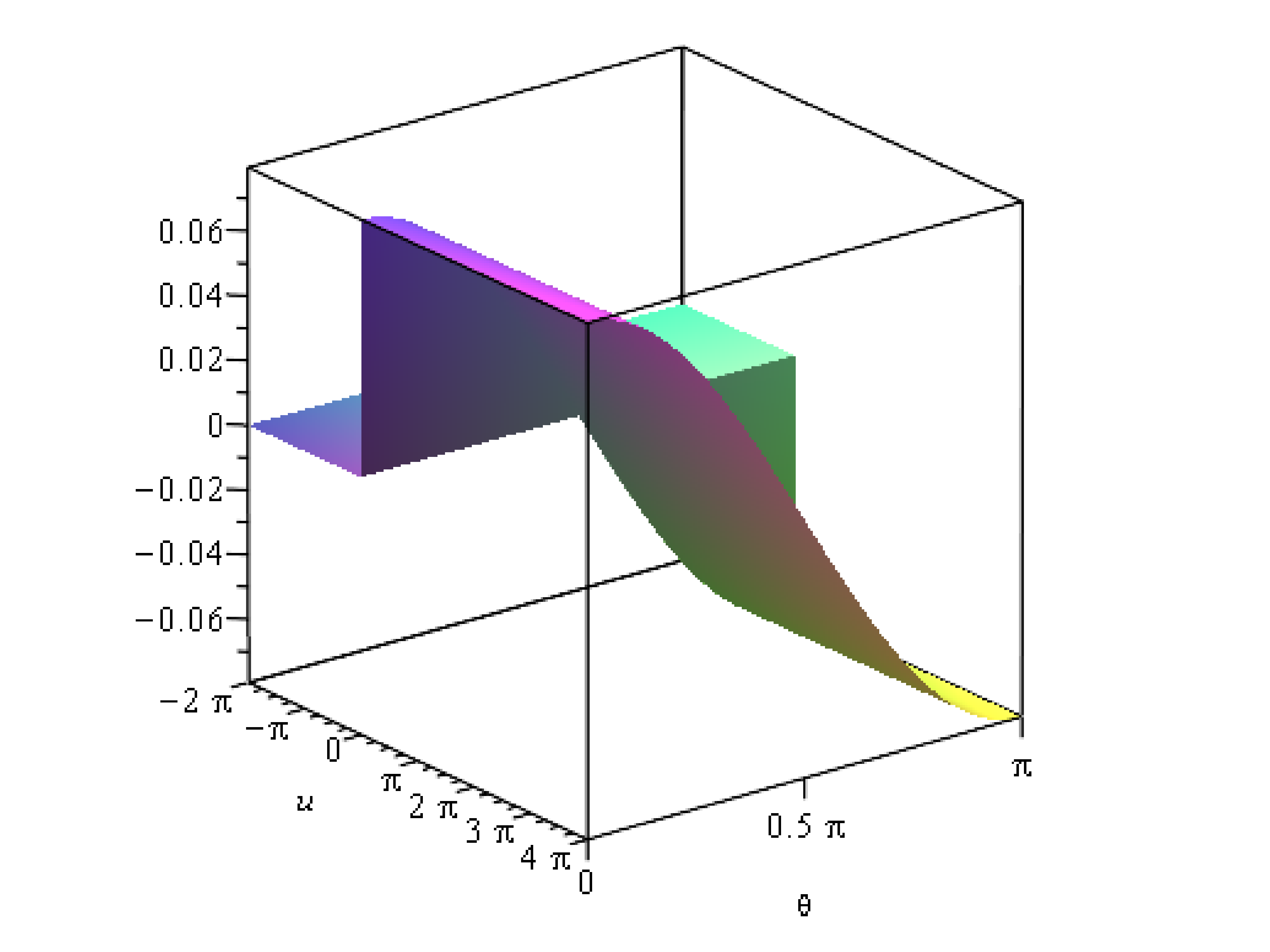}
        \end{center}
  \caption{The time evolution of the E-mode memory ( $q=1,V=1/2$ )}
   \label{F.1}
  \end{center}
\end{figure}
We plotted this in Fig.$\ref{F.1}$. 
This solution is consistent with that in \cite{0264-9381-31-20-205003}.
Using this background electric field, we can deduce the perturbed B-mode electric field as
\bna
&& \lambda\ep^{AB}E_{A:B}^{(1)} = \frac{qVDm\lambda}{8\pi}\cos{\theta} \sin{\left\{ \frac{m}{2} \left( 2\,r+u \right)  \right\} } H(u) \no \\
&& \hspace{2cm}- \frac{qVD\lambda}{4 \pi}\cos{\theta} \cos \left\{ \frac{1}{2}\,m \left( 2\,r+u \right)  \right\} \delta( u ) .
\label{Bmem}
\ena 
Integration with respect to the null time coordinate $u$ gives rise to the time evolution of the B-mode electromagnetic memory, that is,
\bna
&&\int du \,\,\lambda\ep^{AB}E_{A:B}^{(1)} \left( r \to \infty \right) \no \\
&& \hspace{5mm} = - \frac{1}{4}\frac{qVD\lambda}{\pi}\cos{\left( \theta \right)} \cos{ \left\{ \frac{m}{2} 
\left( 2 r + u \right)  \right\}} H( u ) \ .
\label{main}
\ena
Here, we set $r$ dependence to zero.
So, we find that there exists the nontrivial and dynamical B-mode electromagnetic memory in axion electrodynamics. 
In Fig.$\ref{F.2}$, we plotted the time evolution of the B-mode memory.
Since axion is oscillating, the B-mode memory also appears oscillating.

Now, we estimate the order of magnitude of the B-mode electromagnetic memory. 
The strength of this B-mode memory depends on the amplitude of axion oscillation $D$, 
a coupling constant $\lambda$, the charge $e$, the velocity $V$ and the distance from the source to us. 
The amplitude $D$ is determined by the energy density of axion in the halo of our Galaxy and the axion mass. 
From observations, we know the density of axion dark matter is about $0.3 \,{\rm GeV/cm^{3}}$.
Assuming the axion mass to be $m=10^{-21}$ eV, we can estimate the amplitude $D$ as $2.1\times 10^{18}\ {\rm eV}$. Suppose the 
 coupling constant is given by $\lambda = 10^{-12}\ {\rm GeV^{-1}}$. The amplitude of the B-mode memory in ($\ref{main}$) is smaller than the E-mode memory of this source by $D\lambda \sim 10^{-3}$. The coefficient of the first term in ($\ref{main}$) is 
 $\frac{qVD\lambda}{4\pi}\sim 3.2\times 10^{14}$. 
 Here, we set $q = 1\ {\rm C}$, $V=c$, where $c$ is the velocity of light.
 Note that $D\propto 1/m$. Hence, the amplitude of the B-mode is proportional to the inverse of the axion mass.
 Since this quantity is the angular component of the field, we must convert it into that in the Cartesian coordinate. Let us take an example, which is the observation of thunder. This phenomenon is the electric discharge of a few Coulombs at least, and we can usually observe it in a few kilometers.
 Then, in this situation, we can estimate the B-mode memory as 
 $$
 \frac{qVD\lambda}{4\pi r} \sim 10^{6} \left(\frac{10^{-21}{\rm eV}}{m}\right)\left(\frac{\lambda}{10^{-12}\ {\rm GeV}^{-1}}\right)
 \left(\frac{ 1{\rm km}}{r}\right) \left(\frac{ q}{1{\rm C}}\right)\left(\frac{ V}{c}\right) \ {\rm eV} 
 \ .
 $$ 
Here, $r$ is the distance from the thunder in Cartesian coordinates. Note that the information of $r$, $q$, and $V$ can be extracted from the observation of the E-mode memory.
 Thus, the amplitude of the B-mode memory tells us the information of the axion coupling constant and the mass of the axion, while the oscillation frequency tells us the mass of the axion. For example, the axion on the halo scales is coherently oscillating with the period of oscillation, 
$$
 {\rm a \ few \ years} \times \left(\frac{10^{-21}{\rm eV}}{m}\right) \ .
$$ 
Thus, we can resolve the degeneracy of the model parameters. Although this effect is quite small for the QCD axion with mass around $10^{-6}$ eV, we may be able to detect the B-mode electromagnetic memory for the ultralight axion. 
 If the axion mass is $10^{-21}\ {\rm eV}$, the oscillation period is a few years, while the period becomes a few seconds for $m=10^{-14}\ {\rm eV}$.
 These are also in a detectable range.
 Hence, it would be worth designing the axion detectors using the effect we found.
 
 We should point out that the magnitude of the B-mode electromagnetic memory is limited by the stringent constraint on the coupling constant. 
However, as for the Chern-Simons gravity, we do not have such a strong constraint.
Therefore, it is worth studying the axion-induced memory in the Chern-Simons gravity coupled with the axion.

\section{Conclusion}

We studied the memory effect in axion electrodynamics.
We have assumed the coherently oscillating axion dark matter exists in the halo of our galaxy. 
We employed the perturbation method with a coupling constant as an expansion parameter. 
We explicitly demonstrated that there can be the B-mode memory in axion electrodynamics. 
The parity violation was essential for obtaining the B-mode memory. 
We have also estimated the magnitude of the B-mode electromagnetic memory to see the detectability.
Although the amplitude is quite small, it is not impossible to detect the axion dark matter through the effect we have found in this work. 

We can extend our analysis to axion Chern-Simons gravity. Since the theory allows us to have the parity violation, there should be the B-mode gravitational wave memory. If the dark matter is an axion, an oscillating axion is ubiquitous in the Universe.
By observing these B-mode gravitational waves, we can obtain the information of the axion dark matter.
Moreover, the amplitude of oscillation may carry the information of dark energy\cite{Aoki:2016mtn,Aoki:2016kwl,Aoki:2017ehb}. 
Thus, the observation of the B-mode gravitational memory would open a new window for exploring the Universe. We leave this issue for future work.

\begin{acknowledgments}
We would like to thank A. Aoki, A. Ito, and K. Tomoda for useful discussions. D. Y. was supported by Grant-in-Aid for JSPS Research Fellow and JSPS KAKENHI Grant No 17J00490. J. S. was supported by JSPS KAKENHI Grant No. 17H02894, 17K18778, and MEXT KAKENHI Grant No. 15H05895, 17H06359.
\end{acknowledgments}



\bibliography{MEAE}

\end{document}